\documentclass[twocolumn,superscriptaddress]{revtex4}
\usepackage{graphicx}
\usepackage{epsfig}
\usepackage{amsmath}
\usepackage{amsfonts,amsbsy}
\usepackage{amssymb}
\usepackage{hyperref}
\usepackage{bm}
\usepackage{color}
\usepackage{enumerate}
\usepackage{verbatim}

\def\be{\begin{equation}}
\def\ee{\end{equation}}
\def\bea{\begin{eqnarray}}
\def\eea{\end{eqnarray}}

\begin{document}
\title{Investigations into the characteristics and influences of nonequilibrium evolution}
\author{Xiaobing Li}
\affiliation{Key Laboratory of Quark and Lepton Physics (MOE) and Institute of Particle Physics,\\
Central China Normal University, Wuhan 430079, China}
\author{Mingmei Xu}
\email{xumm@ccnu.edu.cn}
\affiliation{Key Laboratory of Quark and Lepton Physics (MOE) and Institute of Particle Physics,\\
Central China Normal University, Wuhan 430079, China}
\author{Yanhua Zhang}
\affiliation{Department of physics and electronic engineering, Yuncheng University, Yuncheng 044000, China}
\author{Zhiming Li}
\affiliation{Key Laboratory of Quark and Lepton Physics (MOE) and Institute of Particle Physics,\\
Central China Normal University, Wuhan 430079, China}
\author{Yu Zhou}
\affiliation{Department of Mathematics, University of California, Los Angeles, California 90095, USA}
\author{Jinghua Fu}
\affiliation{Key Laboratory of Quark and Lepton Physics (MOE) and Institute of Particle Physics,\\
Central China Normal University, Wuhan 430079, China}
\author{Yuanfang Wu}
\email{wuyf@ccnu.edu.cn}
\affiliation{Key Laboratory of Quark and Lepton Physics (MOE) and Institute of Particle Physics,\\
Central China Normal University, Wuhan 430079, China}
\date{\today}
\begin{abstract}
In order to estimate qualitatively the influence of nonequilibrium evolution in relativistic heavy ion collisions, we use the three dimensional Ising model with Metropolis algorithm to study the evolution from nonequilibrium to equilibrium on the phase boundary. The evolution of order parameter approaches its equilibrium value exponentially, the same as that given by Langevin equation. The average relaxation time is defined which is demonstrated to well represent the relaxation time in dynamical equations. It is shown that the average relaxation time at critical temperature diverges as the $z$th power of system size. The third and the fourth cumulants of order parameter during the nonequilibrium evolution could be either positive or negative, depending on the observation time, consistent with dynamical models at $T>T_{\rm c}$. It is found that the nonequilibrium evolution at $T>T_{\rm c}$  lasts very short, and the influence is  weaker than that at $T<T_{\rm c}$. Those qualitative features are instructive to determine experimentally the critical point and the phase boundary of QCD.
\end{abstract}

\maketitle

\section{Introduction}
One goal of current relativistic heavy ion collision experiments is to map the QCD phase diagram as a function of temperature $T$ and baryon chemical potential $\mu_{\rm B}$. QCD phase transition is a nonperturbative process whose analytical solutions are difficult. Lattice QCD is restricted to vanishing or small baryon chemical potential, where the hadron phase smoothly transitions to QGP phase~\cite{lab1}. QCD-based models predict the phase transition is first order at low $T$ and high $\mu_{\rm B}$~\cite{first}. The first order phase transition line ends at a critical point~\cite{cp}. The critical sensitive observables are high cumulants of conserved charges ~\cite{high-cumu,high-cumu2}.

Currently, at the RHIC beam energy scan, non-monotonous behavior of the fourth cumulants (kurtosis) of net protons has been observed~\cite{STAR}. While, the third cumulants (skewness) of net protons are always negative below the Poisson baseline~\cite{STAR}, in contrast to the positive values obtained by the equilibrium expectations~\cite{Stephanov-1999,Stephanov-2011,Asakawa}. The negative values may be caused by nonequilibrium effects~\cite{Mukherjee-2015, Nahrgang-2019, Song-2019, Rajagopal-2019}.

Usually, equilibrium is assumed in lattice QCD and various QCD-based models~\cite{Mukherjee-2015,Nahrgang-2019,Berdnikov-2000}. However, in reality, nonequilibrium is inevitable. 
The fireball formed in a heavy ion collision only spends a limited time. The initial state of the fireball may reach, or not reach thermal equilibrium~\cite{PBM-nature,equi-exp}. 

If the collision energy is high enough and the initial state reaches the temperature needed for the deconfinement phase transition, the fireball transforms to a deconfined phase consisting of quarks and gluons. In this case, the QGP phase is formed and the system reaches equilibrium~\cite{equi-exp}. It is also possible that local equilibrium is reached and QGP is formed in droplets. 
Another possibility is that some collisions in heavy ion experiments do not experience the phase transition, and the system does not reach equilibrium. Therefore, equilibrium, local equilibrium and nonequilibrium are three possible initial states. 

In addition, at the critical region, even if the initial state reaches equilibrium, the critical slowing down prevents the system from equilibrium~\cite{Berdnikov-2000}.

For the statistics of equilibrium state, phase transition is due to spontaneous symmetry breaking. Systems sharing the same symmetry have the same critical exponents, and belong to a universality class. Critical points of both QCD and 3D Ising model belong to the Z(2) symmetry group~\cite{univers-1,lab5,univers-3,univers-4,Z2-1}. So in order to learn the critical phenomena of QCD, people often refer to the 3D Ising model. 

The singular part of the QCD free energy density can be obtained by means of the scaling function of the Ising model~\cite{Mukherjee-2015, scaling}. The mapping of the Ising phase diagram to that of the QCD is also constructed~\cite{Stephanov-2011}. The mapping makes the temperature axis of the Ising model tangent to the first order phase transition line of the QCD, ensuring the coincidence of critical points and approximate coincidence of first order phase transition lines~\cite{mapping2}. High cumulants of conserved charges in relativistic heavy ion collisions are approximately corresponding to those of magnetization of 3D Ising model.

As far as nonequilibrium evolution is concerned, there are no ready theories at present. The common feature of nonequilibrium evolution is that relaxation time at critical point diverges by the power of dynamic exponent. The dynamic exponent is system and algorithm dependent. 

Dynamical evolution equations, e.g. Langevin dynamics~\cite{Song-2019,Berdnikov-2000} and various relaxational models~\cite{Mukherjee-2015, Nahrgang-2019}, are usually used to estimate the effects of nonequilibrium in the QCD critical region. However, the solutions of dynamical evolution equations are restricted to the region of crossover. 

As we know, the numerical simulation of the Ising model with Metropolis algorithm is suitable for studying nonequilibrium evolution~\cite{lab20}\cite{lab26}. The relaxation processes can be easily realized on the whole phase boundary. We find that the nonequilibrium evolution of the order parameter in the Ising model indeed approaches exponentially to its steady value, the same as that of the Langevin equation~\cite{Landau}. The average relaxation time at critical temperature diverges as the $z$th power of system size, the same as relaxation time in dynamical equations~\cite{lab33}. Moreover, the third and fourth cumulants of order parameter on the crossover side oscillate around zero, and then converge to their equilibrium values. The sign of the third cumulants can be negative, consistent with those obtained from refs.~\cite{Mukherjee-2015,Nahrgang-2019}.

In this paper, we first demonstrate the characteristics of nonequilibrium evolution and introduce a time scale of nonequilibrium evolution. Then the influences of nonequilibrium evolution on observables are presented. Section II gives a short introduction to Ising model and Metropolis algorithm. In Section III, the dependences of average relaxation time on temperature, system size and initial configuration are investigated. Characteristics of nonequilibrium evolution are presented. Section IV presents the time evolution of cumulants of order parameter. Influences of nonequilibrium on observables are discussed. A summary is given in section V.

\section{Ising Model and Metropolis algorithm}

The 3D Ising model considers a three dimensional cubic lattice composed of $N=L^3$ sites, where $L$ is called the system size. Every site $i$ is occupied by a spin, $s_i$. The spins can be in one of two states, either spin-up, $s_{i}=+1$, or spin-down, $s_{i}=-1$. The state of the system can be represented by a series of spins, i.e.
\begin{equation}
\{s_{1},  s_{2}, \cdots, s_{N}\}.
\end{equation}
A shorthand notation $\{s_i\}$ is used in the following. 

The spins at positions $i$ and $j$ interact with one another. For a pair of parallel spins we assign an interaction energy of $-J$, while for a pair of anti-parallel spins we assign an interaction energy of $+J$. Only interactions with the nearest neighbors are considered.

The spins also interact with an external magnetic field $H$. The total energy of a system of $N$ spins with constant nearest-neighbor interactions $J$ placed in a uniform external field $H$ is 
\begin{equation}
E_{\lbrace s_{i}\rbrace}=-J\sum_{\langle ij\rangle}s_{i}s_{j}-H\sum_{i=1}^N s_{i},
\end{equation}
where the notation $\langle ij\rangle$ restricts the sum to run over all the nearest neighbor spins. 

Then the partition function is
\begin{equation}
Z(T,H)=\sum_{\{s_{i}\}}{\rm exp}(- E_{\lbrace s_{i}\rbrace}/k_{\rm B}T),        
\end{equation}
where $k_{\rm B}$ is Boltzmann's constant. The free energy is evaluated by
\begin{equation}
F(T,H)=-k_{\rm B}T{\rm ln}Z.
\end{equation}
The average total magnetization is
\begin{equation}
M=-(\frac{\partial F}{\partial H})_{T}=\langle \sum_{i=1}^N s_{i} \rangle,
\end{equation}
and the per-spin magnetization is 
\begin{equation}
m=\frac{1}{N} \sum_{i=1}^N s_{i}.
\end{equation} 
 
We here focus on the behavior of the Ising model as a function of temperature $T$, which defines an energy scale $k_{\rm B}T$. For temperatures $k_{\rm B}T\ll J$, the spin-spin interactions are relatively strong, so that the spins tend to align with one another, and $\vert m\vert$ is close to unity. This is an ordered phase. For temperatures $k_{\rm B}T\gg J$, the spin-spin interactions are relatively weak, so that the spins are effectively non-interacting and point up and down randomly, and $\vert m\vert$ is close to zero. This is a disordered phase. A phase transition from a high-temperature disordered phase to a low-temperature ordered phase is anticipated which is continuous and called the Curie point. The Curie point is the critical point on the plane of variables $T$ and $H$. The temperature of the critical point is denoted by $T_{\rm c}$.

The value of $T_{\rm c}$ of the two dimensional Ising model was exactly calculated by Onsager in 1944 \cite{lab28}. $T_{\rm c}$ of the 3D Ising model is estimated by the finite size scaling theory and analysis of magnetization distribution.  $K_{\rm c}= J/k_{\rm B}T_{\rm c}= 0.2216544$ is obtained~\cite{lab31}, which agrees very well with the results obtained from the renormalization group theory \cite{lab32}. Usually, $J$ and $k_{\rm B}$ are set to 1, so $T_{\rm c}= 4.51$, which is used in the following.

Metropolis algorithm was introduced by Nicolas Metropolis and his collaborators in their paper in 1953 \cite{lab27}. The main steps of Metropolis algorithm in the 3D Ising model are as follows.
\begin{enumerate}[(1)]
\item Set the temperature $T$ and the number of sites $N$.

\item Generate the initial configuration. 

Two kinds of initial configuration are usually used. One is random configuration with all spins pointing randomly up or down, while the other one is polarized configuration with all spins pointing in the same direction.

\item Test a single spin whether flips or not. 

Whether a spin is flipped depends on the acceptance probability $A({\pmb u}\rightarrow {\pmb v})$, which is given by 
\begin{equation}
A({\pmb u}\rightarrow {\pmb v})=\left\{\begin{array}{ll}
e^{-(E_{\pmb v}-E_{\pmb u})/k_{\rm B}T}&\text{if $E_{\pmb v}-E_{\pmb u}>0$},\\1&\text{otherwise.}\end{array}\right .
\end{equation}
${\pmb u}$ and ${\pmb v}$ represent the state of the system before and after flipping this spin.

If $A({\pmb u}\rightarrow {\pmb v})=1$, the spin is flipped.
  
If $A({\pmb u}\rightarrow {\pmb v})<1$, a random number $r$ ($0<r<1$) is generated. If $A({\pmb u}\rightarrow {\pmb v})>r$, the spin is flipped. Otherwise, the spin keeps its original state.

The testing of one single spin is called a Monte Carlo step.
  
\item  When $N$ Monte Carlo steps are completed, every spin in the lattice has been tested for flipping and we say \textit{one sweep} is completed. In this way, the configuration of the system is updated once a sweep. 

\item After evolving enough sweeps, the magnetization approaches a steady value and the system reaches equilibrium. Thermodynamic quantities are usually measured in the equilibrium state. 
   
\item Change the temperature or the system size and repeat the above steps. 
\end{enumerate}

Metropolis algorithm flips one single spin at a step to get a new state. The Wolff algorithm flips one cluster at a step. While different algorithms give the same equilibrium state, the ways in which the system comes to equilibrium are different. In this sense, one algorithm acts as one dynamic, and realizes one relaxation process from nonequilibrium to equilibrium.

The dynamic exponent $z$ of the Metropolis algorithm and the Wolff algorithm for 3D Ising model is 2.02 and 0.33, respectively~\cite{lab17}. Due to the large dynamic exponent, the Metropolis algorithm is more suitable for studying relaxation processes~\cite{lab20}\cite{lab26}.

By using Metropolis algorithm, we simulate the evolution of the 3D Ising model from nonequilibrium to equilibrium at vanished external field and obtain samples for the next calculations. 

\section{Characteristics of nonequilibrium evolution}

Starting from an initial configuration, the Ising system can evolve to an equilibrium state of a given temperature spontaneously. The evolution before reaching equilibrium is called the relaxation process, or \textit{nonequilibrium evolution}.

Starting from random configurations, we simulate 5000 evolution processes for each temperature at a system size $L=60$. Fig.~1(a)-(d) show the time evolution of $\vert m\vert$ (the absolute value of $m$ is used because the sign of the magnetization is random at $H=0$) at four temperatures, i.e. $T/T_c=0.93$,  0.99, 1.00 and 1.03, respectively. The first two temperatures are representative values on the first order phase transition line near the critical point. $T/T_c=1.00$ represents the critical point. $T/T_c=1.03$ is a representative value on the crossover side. The horizontal axis is time which is defined by the number of sweeps introduced in the fourth step of Metropolis algorithm in section II. The red curve and the blue curve are the results of two evolution processes randomly selected from the sample. 

\begin{figure*}[ht]
\centering
\includegraphics[width=0.95\textwidth]{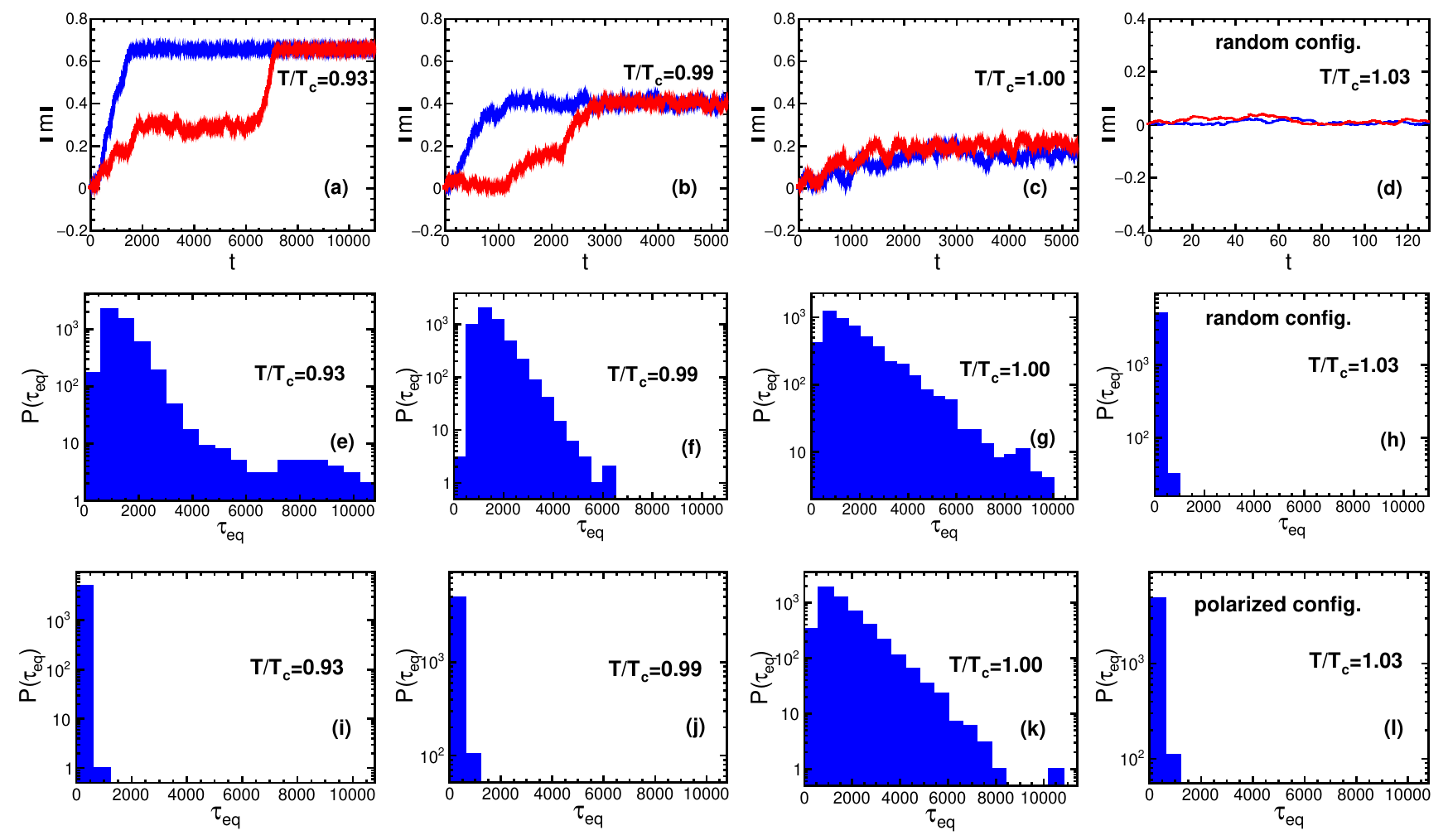}
\caption{(a)-(d): The evolution of $\vert m\vert$ with time at $L=60$ for random initial configuration at temperatures $T/T_{\rm c}=0.93$,  0.99, 1.00 and 1.03, respectively. The blue curve and the red curve represent two evolution processes randomly selected from the sample. (e)-(h): The distribution of relaxation time $\tau_{\rm eq}$ for random initial configuration at four temperatures. (i)-(l): The distribution of relaxation time $\tau_{\rm eq}$ for polarized initial configuration at four temperatures.}
\end{figure*}

In Fig.~1(a), $\vert m\vert$ shows an increasing trend at the beginning, and then gets close to a steady value. After that, $\vert m\vert$ fluctuates slightly around the steady value. The steady value represents an equilibrium state, and the approaching to the steady value represents the relaxation to equilibrium. 

We use $\mu$ to denote the steady value, i.e. the equilibrium expectation. When the difference with the equilibrium expectation is considerably larger than the root of the variance at equilibrium $\sigma=\langle(x-\mu)^2\rangle^{1/2}$, the system is far from equilibrium. 

The period of time from nonequilibrium to equilibrium is called the relaxation time~\cite{lab17}\cite{lab18}. In our simulation, relaxation time of the $i^{\rm th}$ evolution process $\tau_{\rm eq}^{i}$ is estimated by the time when the value of $\vert m\vert$ enters the interval ($\mu -\sigma$, $\mu + \sigma$), i.e. the band of thermal fluctuations around the equilibrium expectation. 

Relaxation time of the $i^{\rm th}$ process defined above represents the number of iterations needed to achieve equilibrium. One iteration is counted after each of the spins is examined to flip or not by given dynamics. We count the number of iterations, which is an integer. The number of iterations presents the steps, i.e. time, that the system needs to achieve equilibrium.


In Fig.~1(a) the steady values of the red curve and the blue curve are the same, but relaxation time of the red curve is much longer than the blue curve. The difference of relaxation time between the red curve and the blue curve is significant at low temperature as Fig.~1(a) shows and seems diminishing at high temperature as Fig.~1(b)-(d) show. In order to show the difference of relaxation time of different evolution processes, the distributions of $\tau_{\rm eq}$ are plotted in Fig.~1(e)-(h). For the sake of comparison, the horizontal ordinate of the four figures are set to the same.  

At the temperature $T/T_{\rm c}=0.93$, the distribution of $\tau_{\rm eq}$ has a long tail, as shown in Fig.~1(e). The long tail means there are a fraction of evolution processes whose relaxation time is very long. When the temperature gets closer to $T_{\rm c}$, the distribution gets narrow, as shown in Fig.~1(f). At the critical temperature, the distribution gets wide again, as Fig.~1(g) shows. The number of systems with a long relaxation time increases at the critical temperature. On the crossover side, i.e. $T/T_{\rm c}=1.03$, the width of the distribution is the smallest and the distribution is concentrated at very short relaxation time, as Fig.~1(h) shows. 


To quantify the relaxation time of different evolution processes, we define \textit{the average relaxation time} as
\begin{equation}
\bar \tau_{\rm eq}=\frac{1}{n}\sum_{i=1}^{n} \tau_{\rm eq}^{i},
\end{equation}
where $n$ is the total number of evolution processes,  $\tau_{\rm eq}^{i}$ is relaxation time of the $i^{\rm th}$ evolution process. At $\bar \tau_{\rm eq}$, not all systems are at equilibrium. There is still a proportion of systems in nonequilibrium state. $\bar\tau_{\rm eq}$ should represent the relaxation time $\tau_{\rm eq}^{\rm dyn}$ in dynamical equations. 
 
Generally, the relaxation time depends on the mechanism of dynamic process, the system size, temperature, initial configurations, and so on. In Fig.~2 we systematically illustrate how the average relaxation time varies with temperature, the system size and the initial configuration. 

\begin{figure}[t]
\begin{minipage}{0.5\linewidth}
\centerline{\includegraphics[width=1\textwidth]{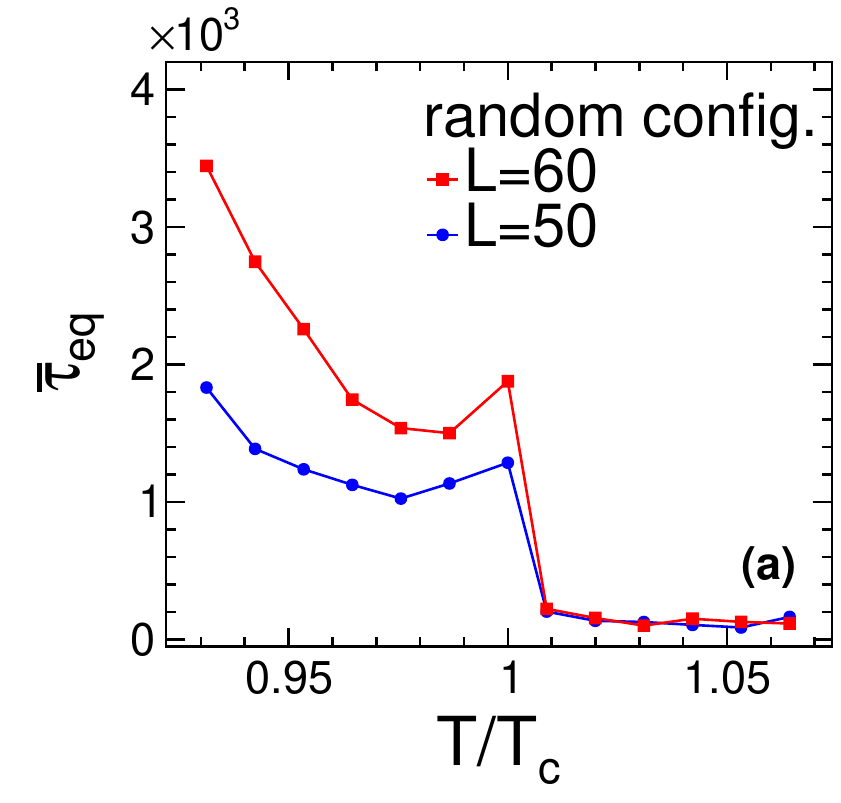}}
\end{minipage}
\begin{minipage}{0.48\linewidth}
\centerline{\includegraphics[width=1\textwidth]{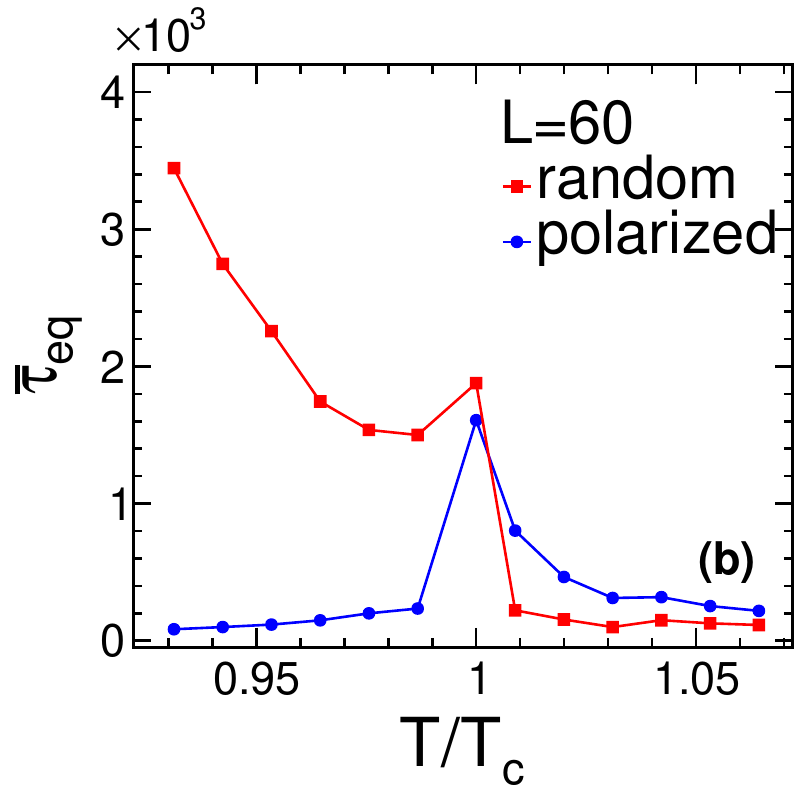}}
\end{minipage}
\caption{Average relaxation time $\bar \tau_{\rm eq}$ as a function of temperature (a) at system sizes $L=50$ (blue circles) and 60 (red squares) starting from random initial configuration; (b) for random initial configurations (red squares) and polarized initial configurations (blue circles) at a fixed system size $L=60$. }
\end{figure}

\begin{figure}[]
\centering
\includegraphics[width=0.35\textwidth]{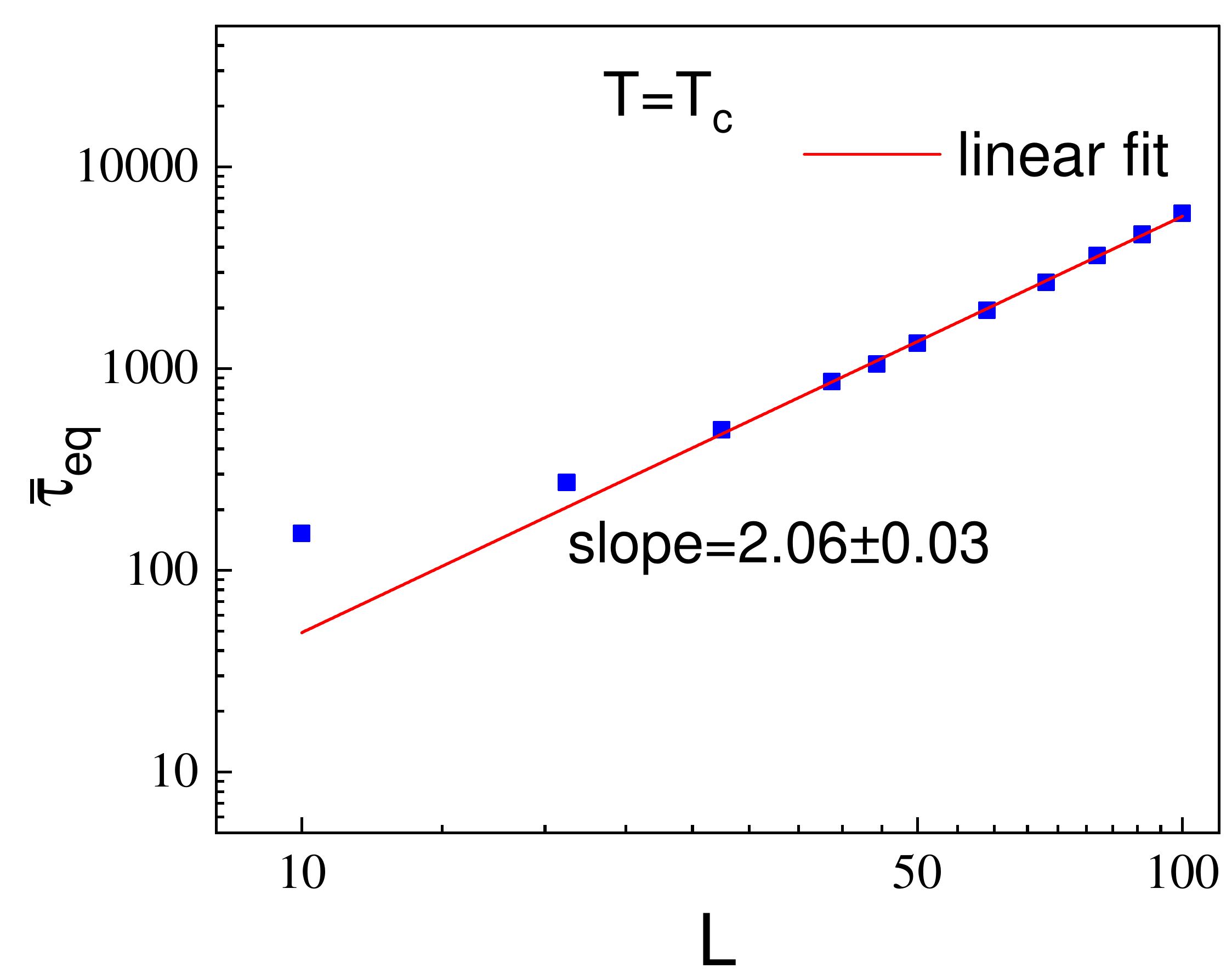}
\caption{The finite size scaling of average relaxation time at critical temperature. The straight line is a linear fit.}
\end{figure}

As Fig.~2(a) shows, in the neighborhood of $T_{\rm c}$, the average relaxation time has a peak which increases with the system size. Fig.~2(b) also shows peaks around the critical temperature no matter what initial configuration is. 

Due to critical slowing down, the relaxation time near $T_{\rm c}$ for an infinite system size is expected to diverge as~\cite{lab33}
\begin{equation}
\tau_{\rm eq}^{\rm dyn}\propto \xi^z \propto |T-T_{\rm c}|^{-z\nu},
\end{equation}
where $\nu$ is the critical exponent of correlation length. $z$ is the dynamic exponent, which governs the dynamic universality class~\cite{1977}. For a finite system size, $\xi\sim L$, Eq. (9) reads
\begin{equation}
\tau_{\rm eq}^{\rm dyn}\propto L^z .
\end{equation}
This gives the power-law behavior of the relaxation time at $T_{\rm c}$. 

In order to test the power-law behavior of $\bar{\tau}_{\rm eq}$, the log-log plot of $\bar{\tau}_{\rm eq}$ versus system size is presented in Fig.~3. Its linear region is well fitted by a straight line with slope equal to 2.06$\pm$0.03, consistent with refs.~\cite{lab17,z-ref1,z-ref2}. $\bar{\tau}_{\rm eq}$ of critical temperature indeed diverges as the z-th power of system size, the same as $\tau_{\rm eq}^{\rm dyn}$. 

We also examine the dependence of the standard error and the relative standard error of $\overline{\tau}_{\rm eq}$ on system size. The standard error increases with system size, and the relative standard error is almost constant. This indicates a violation of self-averaging at the critical temperature~\cite{self-averaging}. The violation of self-averaging appears to be a common property at criticality. 

In Fig.~2(a), for a fixed system size, $\overline{\tau}_{\rm eq}$ shows a decreasing trend with increasing temperature on both sides of $T_{\rm c}$ except the neighborhood of $T_{\rm c}$. This trend originates from that the acceptance probability $A({\pmb u}\rightarrow {\pmb v})$ in Eq.~(7) is an increasing function of  temperature $T$. Higher temperature, higher acceptance probability and then shorter relaxation time.  

For $T>T_{\rm c}$, the average relaxation time is weakly dependent on the system size and the initial configuration, as demonstrated in both Fig.~2(a) and (b). No matter what the system size is and what the initial configuration is, the average relaxation time is always short and has a trend towards zero.  This is due to larger acceptance probability for higher temperature. With larger acceptance probability, it is easier for the system to change from one configuration to another and hence the relaxation time is extremely short. 

In contrary, for $T<T_{\rm c}$, the average relaxation time has strong dependences on system size and initial configuration. First, Fig.~2(a) demonstrates that the larger system size, the larger average relaxation time. It is natural that larger system needs more time to get equilibrium. Second, Fig.~2(b) demonstrates that the average relaxation time of random configurations is much longer than polarized configurations. This is because equilibrium state of $T<T_{\rm c}$ is close to the ordered phase. Polarized configurations are ordered, and random configurations are disordered. So polarized initial configuration evolves to an ordered state faster than that of random initial configuration. 

The effect of initial configuration is also demonstrated by the distribution of $\tau_{\rm eq}$ in the second row (for random initial configuration) and the third row (for polarized initial configuration) in Fig.~1, respectively. Wide distributions which extend to large $\tau_{\rm eq}$ result in large $\bar \tau_{\rm eq}$, while narrow distributions which are concentrated at small $\tau_{\rm eq}$ result in small $\bar \tau_{\rm eq}$.

It is natural that the further the initial configuration deviates from the equilibrium state, the longer the relaxation time is. Therefore, on the left side of $T_{\rm c}$, random configuration has longer average relaxation time, while on the right side of $T_{\rm c}$ polarized configuration has longer average relaxation time, as demonstrated in Fig.~2(b).    

In order to observe the maximum influences of nonequilibrium, initial configurations that have longer nonequilibrium evolution are selected, i.e. random initial configuration for $T\leqslant T_{\rm c}$ and polarized initial configuration for $T>T_{\rm c}$ are used in the following. 

\section{Influences of nonequilibrium evolution}

In this section the influences of nonequilibrium evolution on observables are demonstrated in two ways, i.e. how observables vary with evolution time and with temperature.  

It is shown in last section that the distribution of relaxation time varies with temperature. Even for a fixed temperature and a fixed system size, relaxation time of two simulations is likely to be different, as the red curve and the blue curve show in Fig.~1(a). If measurements are made at a time when not all systems reach thermal equilibrium, there is a certain proportion of nonequilibrium systems in the sample, very similar to the initial states of relativistic heavy ion collisions. 

High order cumulants of order parameter are sensitive observables in the search of a critical point. Letting $X=\vert m\vert$, $\delta X=\vert m\vert-\langle \vert m\vert \rangle$, the first four cumulants read
\begin{eqnarray}
C_{1}&=& \langle X\rangle,\\
C_{2}&=& \langle(\delta X)^{2}\rangle,\\
C_{3}&=& \langle(\delta X)^{3}\rangle,\\
C_{4}&=& \langle(\delta X)^{4}\rangle-3\langle(\delta X)^{2}\rangle^2.
\end{eqnarray} 
$C_1$ is the mean value of order parameter distribution. $C_2$ is the variance of the distribution. $C_3$, $C_4$ are related to skewness and kurtosis, which can quantify non-Gaussianity of the distribution.
 
\begin{figure*}[htbp]
\centering
\includegraphics[width=1\textwidth]{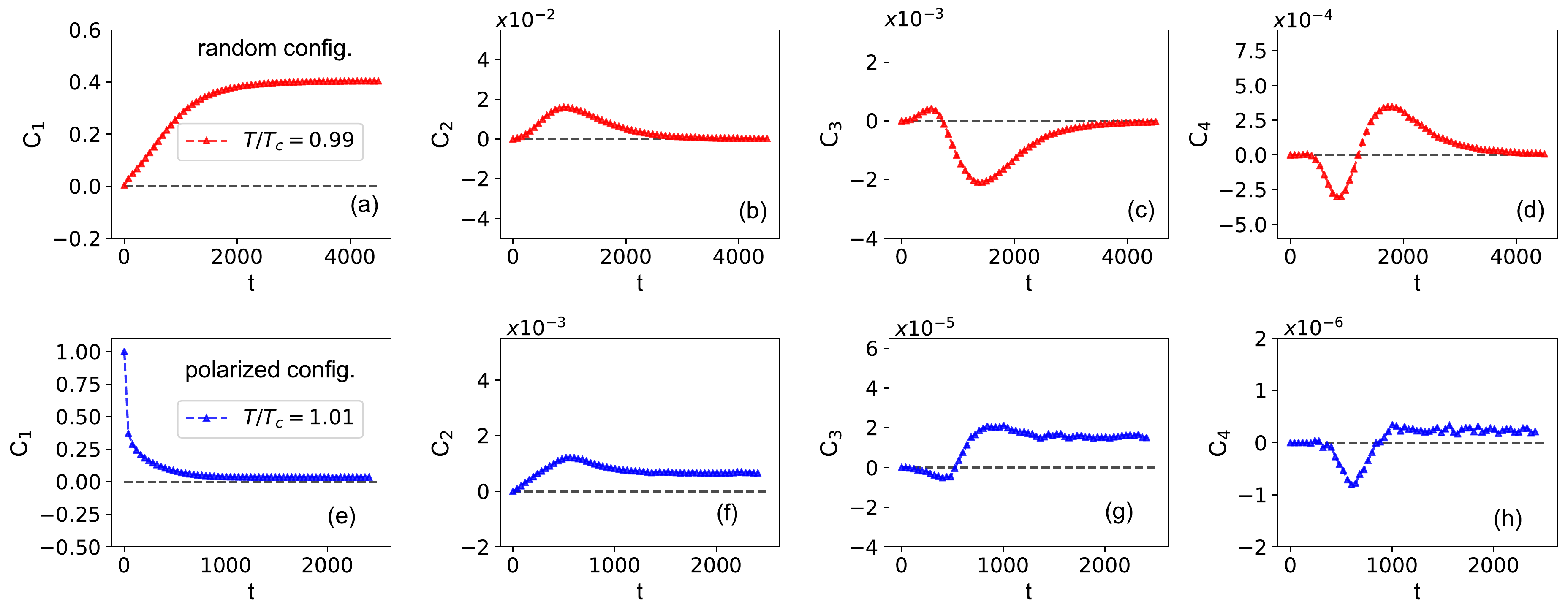}
\caption{The time evolution of cumulants of $\vert m \vert$ at a given system size $L=60$ at $T/T_{\rm c}=0.99$ (the upper row) and 1.01 (the lower row). Random initial configuration is used at $T/T_{\rm c}=0.99$ and polarized initial configuration is used at $T/T_{\rm c}=1.01$. }
\end{figure*}
 
Fig.~4 demonstrates how cumulants vary with the evolution time at two given temperatures. The evolution starts at $t=0$. At $t=0$, $C_{n=1,2,3,4}$ in Fig.~4(a)-(d) are all zero due to random initial configuration, while $C_{1}$ is 1 and $C_{2,3,4}$ are zero due to polarized initial configuration as Fig.~4(e)-(h) show.

The time evolution of $C_1$ in Fig.~4 (a) and (e) demonstrates a similar behavior with Langevin dynamics, i.e. exponentially approaching the equilibrium value which is the solution of a linear  differential equation~\cite{Landau}.  Therefore, the single-spin-flip mechanism in Metropolis algorithm equivalently describes the relaxation in Langevin dynamics.

At the temperature on the first order phase transition line, i.e. $T/T_{\rm c}=0.99$, $C_1$ varies monotonously with time until it approaches a steady value, as Fig.~4(a) shows. The steady value is the equilibrium expectation. In Fig.~4(b), $C_2$ increases first and decreases later, forming a peak during the evolution. In Fig.~4(c) and (d), both $C_3$ and $C_4$ experience oscillations before approaching a steady value. The oscillation results in sign changes during the evolution. The sign of $C_3$ and $C_4$ can be either positive or negative, depending on the evolution time. 

At the temperature on the crossover side, i.e. $T/T_{\rm c}=1.01$, the trend of the nonequilibrium evolution of cumulants is similar to that of $T/T_{\rm c}=0.99$, as Fig.~4(e)-(h) show. $C_1$ varies monotonously with time. $C_2$ experiences nonmonotonic changes before approaching a steady value. $C_3$ also shows sign change during the evolution, similar to Fig.~4(c). The sign is negative at first and then becomes positive, in contrary to Fig.~4(c). $C_4$ in Fig.~4(h) also shows sign change, being negative first and then positive, in similar pattern with Fig.~4(d). The sign of $C_3$ and $C_4$ can be either positive or negative, depending on the evolution time. The sign-change behavior of $C_3$ and $C_4$ on the crossover side is consistent with ref.~\cite{Mukherjee-2015,Nahrgang-2019}.

Apart from the trend, the nonequilibrium evolution of cumulants on both sides of $T_{\rm c}$ has big differences in two aspects. First, the time needed to approach equilibrium expectations for $C_3$ is about 4000 at $T/T_{\rm c}=0.99$ and is less than 2000 at $T/T_{\rm c}=1.01$. The difference is more than twice. That is to say, it is more difficult for systems at $T<T_{\rm c}$ to achieve equilibrium in the same amount of time. Second, the magnitude of the oscillation in $C_3$ and $C_4$ at $T<T_{\rm c}$ is much larger than that at $T>T_{\rm c}$, differing about two orders of magnitude. It means that, when systems on both sides of $T_{\rm c}$ suffer from nonequilibrium effect, systems on the low temperature side deviate further from equilibrium expectations than that on the high temperature side.  



\begin{figure*}[ht]
\centering
\begin{minipage}{0.95\linewidth}
\centerline{\includegraphics[width=1\textwidth]{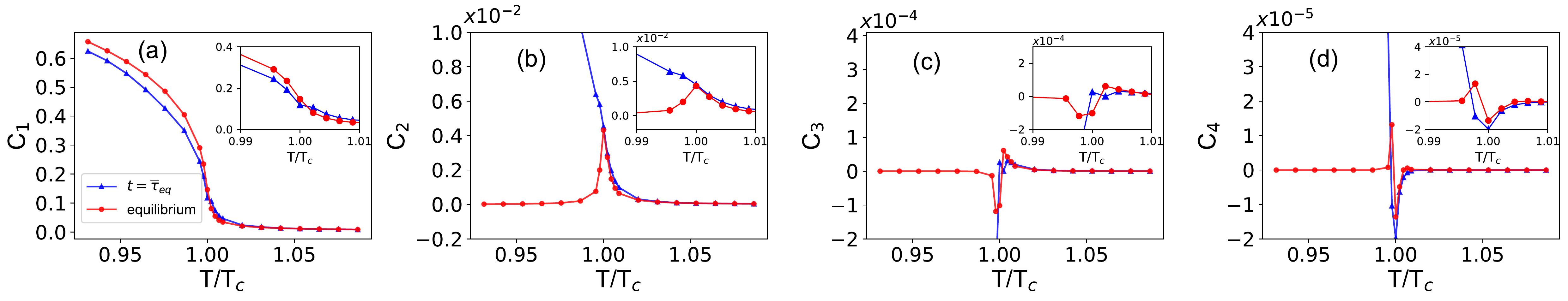}}
\end{minipage}
\qquad
\begin{minipage}{0.95\linewidth}
\centerline{\includegraphics[width=1\textwidth]{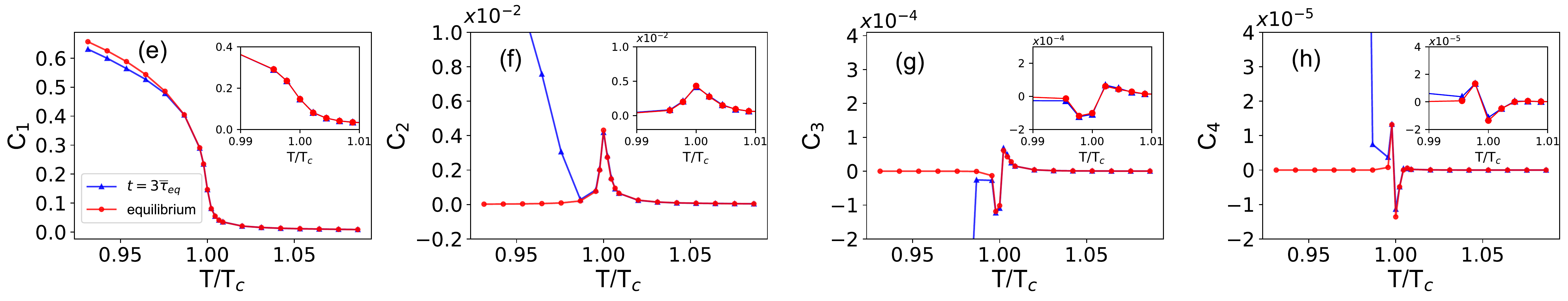}}
\end{minipage}
\caption{Nonequilibrium cumulants of $\vert m \vert$ at a given system size $L=60$ measured at $\bar \tau_{\rm eq}$ (the upper row) and $3 \bar \tau_{\rm eq}$ (the lower row), respectively. Random initial configuration is used for $T\leqslant T_{\rm c}$ and polarized initial configuration is used for $T>T_{\rm c}$.}
\end{figure*}

Since the sign of $C_3$ and $C_4$ during the nonequilibrium evolution depends on the evolution time, two observation times are used as examples to demonstrate in Fig.~5 how observables vary with temperature for a given observation time.
 
In Fig.~5, cumulants in the upper row are measured at $\bar \tau_{\rm eq}$ and that in the lower row are measured at $3\bar \tau_{\rm eq}$. The equilibrium cumulants denoted by red circles are obtained after 250,000 Monte Carlo sweeps, which ensures that all the systems in the sample reach equilibrium. The blue triangles represent the measured value of cumulants at a given observation time during the nonequilibrium evolution. We call them \textit{nonequilibrium cumulants}.

From the upper row of Fig.~5 we can see that the blue triangles are nearly coincident with the red circles at $T>T_{\rm c}$ for all $C_{n=1,2,3,4}$. Near $T_{\rm c}$, the blue triangles start to deviate from the red circles. It is more clearly seen in the insert map of each figure which focuses on the neighbourhood of $T_{\rm c}$. The deviations become larger at $T<T_{\rm c}$. It means that nonequilibrium cumulants observed at $\bar \tau_{\rm eq}$ are not much different from equilibrium cumulants on the high temperature side. The deviations get larger at low temperature. That is to say, low temperature system is more largely affected by the nonequilibrium effect. 

By comparing different orders of $C_n$, we find that the deviation of nonequilibrium cumulants from equilibrium cumulants is modest at low order, especially the first cumulant $C_1$. Cumulants of higher order have greater deviations from equilibrium fluctuations.

As time goes by, more systems are expected to reach equilibrium, and the impact of nonequilibrium should become smaller. The lower row of Fig.~5 which is measured at a later time $3\bar \tau_{\rm eq}$ indeed shows the expected trend, i.e. more points coincide, including all points at $T>T_{\rm c}$ and part of points at $T<T_{\rm c}$. Even in the neighborhood of $T_{\rm c}$ almost no deviation is seen as shown in the insert map of each figure. Deviations only exist for part of points at $T<T_{\rm c}$. 

At an observation time of $3\bar \tau_{\rm eq}$, nearly all points of $C_1$ coincide. Deviations of $C_2$ are modest. Deviations increase drastically in $C_3$ and $C_4$. This proves that relaxation time varies with observables. Relaxation times for higher order cumulants are larger, consistent with expectations~\cite{Mukherjee-2015}. When low order cumulants have reached their equilibrium expectations, high order cumulants may have not reached equilibrium expectations yet. 

Observation time of $3\bar \tau_{\rm eq}$ is enough for most temperatures. For $T\lesssim0.97~T_{\rm c}$ blue points still deviate from red points as shown in Fig.~5(f)(g)(h). This may be caused by the long tail of the $\tau_{\rm eq}$ distribution at low temperature (see Fig.~1(e)). 

\section{Summary and discussion}

In this paper we simulate the nonequilibrium evolution of the 3D Ising model with Metropolis algorithm at zero external magnetic field. The simulation well produces the dynamical features of nonequilibrium evolution on the phase boundary. 

The relaxation time of each evolution is defined as the number of iterations needed to achieve equilibrium. It represents the steps that the system needs to achieve equilibrium. First, the trend of the order parameter approaching a steady value is consistent with Langevin dynamics. It demonstrates that the number of iterations in the numerical simulation is equivalent to evolution time in reality. Second, the average relaxation time at critical temperature indeed diverges as the theoretical prediction Eq. (10). It indicates that $\bar\tau_{\rm eq}$ exactly corresponds to relaxation time $\tau_{\rm eq}^{\rm dyn}$ in dynamical equations. 

It is found that the average relaxation time depends on the system size and temperature. In average, the average relaxation time is short at $T>T_{\rm c}$, almost independent of initial configurations. At $T<T_{\rm c}$, the average relaxation time is much longer than that of $T>T_{\rm c}$ when the initial configuration is far from the equilibrium state. 

The time evolution of the first four cumulants of order parameter is presented. For both $T<T_{\rm c}$ and $T>T_{\rm c}$, $C_3$ and $C_4$ show oscillations and could be either positive or negative, depending on the observation time, which is consistent with the results of dynamical models. 

Since the relaxation time at $T<T_{\rm c}$ is much longer than that at $T>T_{\rm c}$, it is easy to approach equilibrium at the crossover side. While on the line of the first order phase transition, the system is more difficult to achieve equilibrium and is more susceptible to nonequilibrium. Moreover, the influence of nonequilibrium on observables on the crossover side is much weaker than that on the line of the first order phase transition. 

Those qualitative features mentioned above imply that the influence of nonequilibrium evolution in QCD system is most probably negligible at $T>T_{\rm c}$, i.e. on the crossover side, even if there is a fraction of nonequilibrium initial states in relativistic heavy ion collisions. While, the influence should be treated with more caution at $T<T_{\rm c}$, i.e. on the line of the first order phase transition. Indeed, the sign of the third cumulant of order parameter may be negative due to the evolution of nonequilibrium, consistent with the STAR measurements.

By the way, for the nonequilibrium evolution from one equilibrium state to another equilibrium state,  such as a high temperature state is cooled to a low temperature state, we will study in a coming paper.


\section*{Acknowledgement}
We thank Yi Yin, Huichao Song and Shanjin Wu for interesting discussions. This work is supported in part by the NSFC of China under Grant No. U1732271. The numerical simulations have been performed on the GPU cluster in the Nuclear Science Computing Center at Central China Normal University (NSC3).

\providecommand{\href}[2]{#2}\begingroup\raggedright\endgroup

\begin{thebibliography}{10}%
\makeatletter
\providecommand{\hrefCMSnoop }[0]{\@secondoftwo}%
\makeatother
\providecommand{\doi}{\texttt{doi:}\begingroup \urlstyle{tt}\Url}
\bibitem{lab1}Y. Aoki, G. Endrodi, Z. Fodor, S. D. Katz and K. K. Szabo, Nature {\bf 443}, 675 (2006).
 %
 \bibitem{first}M. Asakawa, K. Yazaki, Nucl. Phys. A {\bf 504}, 668 (1989).
 \bibitem{cp}For a review, see C. Blume, Cent. Eur. J. Phys. {\bf 10(6)}, 1245 (2012) (talk given in CPOD 2011 Wuhan).
 \bibitem{high-cumu}M.A. Stephanov, Phys. Rev. Lett. {\bf 102}, 032301 (2009).
 \bibitem{high-cumu2}S. Ejiri, F. Karsch, K. Redlich, Phys. Lett. B {\bf 633}, 275 (2006).
\bibitem{STAR}J. Adam et al. (STAR Collaboration), Phys. Rev. Lett. {\bf 126}, 092301 (2021).
\bibitem{Stephanov-1999}M. A. Stephanov, K. Rajagopal and E. Shuryak, Phys. Rev. D {\bf 60}, 114028 (1999).
\bibitem{Stephanov-2011}M. A. Stephanov, Phys. Rev. Lett. {\bf 107}, 052301 (2011).
\bibitem{Asakawa}M. Asakawa, S. Ejiri and M. Kitazawa, Phys. Rev. Lett. {\bf 103}, 262301 (2009).
\bibitem{Mukherjee-2015}S. Mukherjee, R. Venugopalan and Y. Yin, Phys. Rev. C {\bf 92}, 034912 (2015).
\bibitem{Nahrgang-2019}M. Nahrgang, M. Bluhm, T. Schafer and S. A. Bass, Phys. Rev. D {\bf 99}, 116015 (2019).
\bibitem{Song-2019}S. Wu, Z. Wu and H. Song, Phys. Rev. C {\bf 99}, 064902 (2019).
\bibitem{Rajagopal-2019}K. Rajagopal, G. Ridgway, R. Weller and Y. Yin, Phys. Rev. D {\bf 102}, 094025 (2020).
\bibitem{Berdnikov-2000} B. Berdnikov and K. Rajagopal, Phys. Rev. D {\bf 61}, 105017 (2000).
\bibitem{PBM-nature}P. Braun-Munzinger, J. Stachel, Nature {\bf 448}, 302 (2007).
\bibitem{equi-exp}J. Adams et al. (STAR Collaboration), Nucl. Phys. A {\bf 757}, 102 (2005).  
\bibitem{univers-1}R.D. Pisarski, F. Wilczek, Phys. Rev. D {\bf 29}, 338 (1984).
\bibitem{lab5} 
M. A. Stephanov, K. Rajagopal and E. Shuryak, Phys. Rev. Lett. {\bf 81}, 4816 (1998).
\bibitem{univers-3}M. Asakawa, J. Phys. G: Nucl. Part. Phys. {\bf 36} 064042 (2009).
\bibitem{univers-4}P. de Forcrand, O. Philipsen, Phys. Rev. Lett. {\bf 105} 152001 (2010).
\bibitem{Z2-1}F. Karsch, E. Laermann and Ch. Schmidt, Phys. Lett. B {\bf 520}, 41 (2001).
\bibitem{scaling}B. Friman, F. Karsch, K. Redlich and V. Skokov, Eur. Phys. J. C {\bf 71}, 1694 (2011).
\bibitem{mapping2}A. Bzdak, S. Esumi, V. Koch, J. Liao, M. Stephanov, N. Xu, Physics Reports {\bf 853}, 1 (2020).
%
\bibitem{lab20}Cheng-Wei Liu, A.~Polkovnikov and A.~W.~Sandvik, Phys.\ Rev.\ B {\bf 89}, 054307 (2014).
\bibitem{lab26}M. Acharyya, Phys.\ Rev.\ E {\bf 56}, 2407 (1997).
\bibitem{Landau}Eq. (118.5) in Statistical Physics Part 1, Third edition, L. D. Landau and E. M. Lifshitz, Pergamon Press, 1980.
\bibitem{lab33}
D. T. Son and M. A. Stephanov,
Phys.\ Rev.\ D {\bf 70}, 056001 (2004).
\bibitem{lab28}L. Onsager, Phys. Rev. {\bf 65}, 117 (1944).

\bibitem{lab31}
 A L Talapov,and H W J. Bl$\ddot{o}$te,
J. Phys. A {\bf 29},5727(1996).
\bibitem{lab32}
 G. S. Pawlcy, R. H. Swendsen,D. J. Wallace and K. G. Wilson,
Phys.\ Rev.\ B {\bf 29}, 4030 (1984).
\bibitem{lab27}
N. Metropolis, A. W. Rosenbluth, M. N. Rosenbluth, A. H. Teller, and E. Teller,
J. Chem. Phys. {\bf 21}, 1087 (1953).
\bibitem{lab17} M E J Newman and G T Barkema. Monte carlo methods in statistical physics[M]. Oxford University Press, 1999.
\bibitem{lab18}
Helmut G. K arXiv:0905.1629(2009).

\bibitem{1977}P.C. Hohenberg, B. I. Halperin, Rev. Mod. Phys. {\bf 49}, 435 (1977).
\bibitem{z-ref1}S. Wansleben and D. P. Landau, J. Appl. Phys. {\bf 61}, 3968 (1987).
\bibitem{z-ref2}M. Hasenbusch, Phys. Rev. E {\bf 101}, 022126 (2020).
\bibitem{self-averaging}A. Malakis, N. G. Fytas, Phys. Rev. E {\bf 73}, 016109 (2006).

\end{thebibliography}
\end{document}